\documentclass[preprint,showpacs,preprintnumbers,superscriptaddress,amsmath,amssymb, prl]{revtex4}
\usepackage{graphicx}
\usepackage{dcolumn}
\usepackage{bm}

\begin{document}

\preprint{Ver. \today}

\title{Isotropic Kink and Quasiparticle Excitations in the Three-Dimensional Perovskite Manganite La$_{0.6}$Sr$_{0.4}$MnO$_3$}

\author{Koji~Horiba}
\email{horiba@post.kek.jp}
\affiliation {Photon Factory, Institute of Materials Structure Science, High Energy Accelerator Research Organization (KEK), 1-1 Oho, Tsukuba 305-0801, Japan}

\author{Miho~Kitamura}
\affiliation {Photon Factory, Institute of Materials Structure Science, High Energy Accelerator Research Organization (KEK), 1-1 Oho, Tsukuba 305-0801, Japan}

\author{Kohei~Yoshimatsu}
\affiliation {Photon Factory, Institute of Materials Structure Science, High Energy Accelerator Research Organization (KEK), 1-1 Oho, Tsukuba 305-0801, Japan}
\affiliation {Department of Physics, Graduate School of Science, The University of Tokyo, 7-3-1 Hongo, Bunkyo-ku, Tokyo 113-0033, Japan}

\author{Makoto~Minohara}
\affiliation {Photon Factory, Institute of Materials Structure Science, High Energy Accelerator Research Organization (KEK), 1-1 Oho, Tsukuba 305-0801, Japan}

\author{Enju~Sakai}
\affiliation {Photon Factory, Institute of Materials Structure Science, High Energy Accelerator Research Organization (KEK), 1-1 Oho, Tsukuba 305-0801, Japan}

\author{Masaki~Kobayashi}
\affiliation {Photon Factory, Institute of Materials Structure Science, High Energy Accelerator Research Organization (KEK), 1-1 Oho, Tsukuba 305-0801, Japan}

\author{Atsushi~Fujimori}
\affiliation {Department of Physics, Graduate School of Science, The University of Tokyo, 7-3-1 Hongo, Bunkyo-ku, Tokyo 113-0033, Japan}

\author{Hiroshi~Kumigashira}
\affiliation {Photon Factory, Institute of Materials Structure Science, High Energy Accelerator Research Organization (KEK), 1-1 Oho, Tsukuba 305-0801, Japan}

\date{\today}

\begin{abstract}
In order to reveal many-body interactions in the three-dimensional (3D) perovskite manganite, we have performed an \textit{in situ} angle-resolved photoemission spectroscopy (ARPES) on La$_{0.6}$Sr$_{0.4}$MnO$_3$ (LSMO) and investigated the behaviors of quasiparticles. We observe quasiparticle peaks around the Fermi momentum, both in the electron and the hole bands, and clear kinks throughout the hole Fermi surface in the ARPES band dispersion. The isotropic behavior sharply contrasts to the strong anisotropic quasiparticle excitation observed in layered manganites. These results suggest that polaronic quasiparticles by coupling of the electrons with Jahn-Teller phonons play an important role in the physical properties of the ferromagnetic metallic phase in 3D manganite LSMO.
\end{abstract}


\maketitle

La$_{1-x}$Sr$_x$MnO$_3$ is a typical hole-doped perovskite manganese oxide that has attracted considerable attention because of its unusual physical properties, such as colossal magnetoresistance behavior and the half-metallic nature of its ground state and hence for spintronics applications \cite{CMR}. The exceptional properties originate from the strong interplay between the charge, lattice, orbital, and spin degrees of freedom \cite{MIT}. Consequently, the electronic and magnetic phases of La$_{1-x}$Sr$_x$MnO$_3$ are sensitive to changes in the strength of these mutual couplings, which result in the rich phase diagrams for the manganites. In order to clarify the origin of these properties of La$_{1-x}$Sr$_x$MnO$_3$, we must understand the detailed electronic band structures, especially the interactions of electrons with other degrees of freedom and their anisotropic aspects. Angle-resolved photoemission spectroscopy (ARPES) is a powerful tool for investigating many-body interactions in quasiparticles as functions of binding energy and momentum \cite{Hufner1}. From the line-shape analysis of the ARPES spectra, the momentum-resolved self-energy of quasiparticles, which reflects many-body interactions, has been widely discussed in strongly correlated electron systems, such as high-$T_c$ cuprate superconductors \cite{Valla1, Damascelli} and the surface states of transition metals \cite{Valla2, Schafer, Higashiguchi, Hayashi, Jiang}.

Therefore, ARPES measurements will allow us to clarify the roles of relevant interactions in the physical properties of manganites. Using the momentum-dependent information on the interactions, the remnant Fermi-surface (FS) topology, and the quasiparticle dynamics which are coupled to collective excitations, have been investigated using ARPES measurements. ARPES studies on layered manganites have revealed a pseudogap formation \cite{Dessau} and the existence of quasiparticle excitations with a nodal-antinodal dichotomous characteristic, which is similar to the characteristic feature of high-$T_c$ cuprate superconductors \cite{Mannella1}. The resulting anisotropic kink observed in the layered manganites strongly suggests that the important underlying physics among two-dimensional (2D) strongly correlated oxides.

On the other hand, such a kink has not been observed in the ARPES studies on three-dimensional (3D) perovskite manganite La$_{1-x}$Sr$_x$MnO$_3$ \cite{Shi1, Falub1, Chikamatsu1, Krempasky1, Krempasky2, Tabeno}. In these studies, single-crystal surfaces are prepared using recently developed sophisticated growth techniques for oxide thin films \cite{Horiba1}. Recent progress in the laser molecular beam epitaxy (MBE) technique has enabled the growth of manganite films by controlling their growth process at an atomic level. Using well-defined surfaces of eptaxial films, the band structures of La$_{1-x}$Sr$_x$MnO$_3$ that do not possess cleavable planes have been intensively studied using \textit{in situ} ARPES measurements. Nevertheless, the absence of a kink in 3D manganites has been questioned because relevant many-body interactions in strongly correlated oxides are also expected to be strong in 3D manganites. It should be noted that such a kink has also been observed in the 2D electronic states, which are located at the interface between 3D manganites LaMnO$_3$ and SrMnO$_3$ in their superlattices \cite{Monkman1}.

Thus, in order to better understand the physics of manganites, we must determine whether the kink observed in the layered manganites is inherent in 2D systems or exists in 3D systems, too. However, the lack of information concerning many-body interactions in the electronic structure near the Fermi level ($E_{\rm F}$) in 3D manganites has hindered our understanding of the interaction between electrons with other degrees of freedom in the manganites. In this Letter, we have observed quasiparticle peaks and kinks in the band dispersion of 3D manganite La$_{0.6}$Sr$_{0.4}$MnO$_3$ (LSMO) by a precise and detailed investigation of 3D electronic structure; \textit{in situ} high energy-resolution ARPES measurements using the tunable excitation energy of synchrotron radiation enable us to trace the electronic structures in momentum space in every 3D direction. From their energy and momentum dependence, interactions of electrons with other degrees of freedom are considered to be a possible origin of the kink in LSMO.

Samples of LSMO were grown on the atomically-flat (001) surface of Nb-doped SrTiO$_3$ substrates using a laser MBE method. Growth conditions are detailed in Refs.~\onlinecite{Chikamatsu1} and \onlinecite{Horiba2}. The fabricated films were immediately transferred through an ultrahigh vacuum to the ARPES chamber without exposure to air \cite{Horiba1}. The \textit{in situ} ARPES measurements were carried out at beamline BL-28A of the Photon Factory (PF), KEK,  at the sample temperature of 12~K using circular polarized synchrotron radiation for the excitation light source. The total energy and angular resolutions were set to approximately 20~meV and 0.3$^{\circ}$, respectively. The $E_{\rm F}$ of the samples was calibrated by measuring a gold foil that was electrically connected to the samples.

\begin{figure}
\begin{center}
\includegraphics[width=0.8\linewidth]{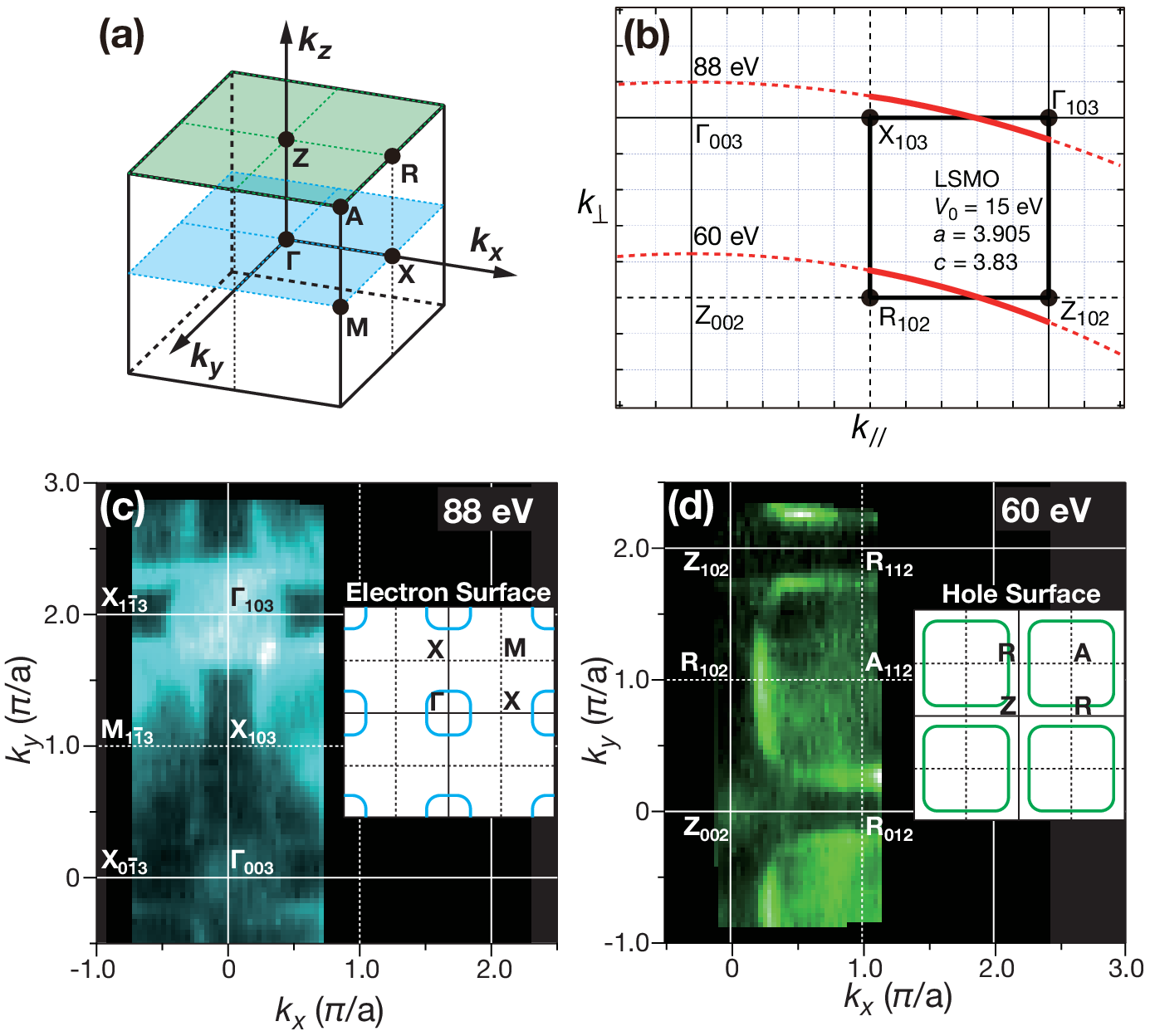}
\end{center}
\caption{(Color online) (a) Brillouin zone for the tetragonal structure of epitaxially strained LSMO films. (b) Measured lines in momentum space for photon energies of 60~eV and 88~eV. (c) FS mapping for photon energies of 88 eV corresponding to the $\Gamma$XM plane and (d) of 60 eV corresponding to the ZRA plane. Insets in Fig.~\ref{figure1} (c) and (d) illustrate the electron pocket around the $\Gamma$ point and the hole pocket around the A point of the Brillouin zone, respectively, which were predicted by band structure calculations \cite{Chikamatsu1, Livesay1}.}
\label{figure1}
\end{figure}

The LSMO films have a tetragonal crystal structure as a result of the epitaxial strain from the SrTiO$_3$ substrates. In order to map FS onto high-symmetry planes in the tetragonal Brillouin zone (Fig.~\ref{figure1} (a)), photon energies of 88 eV for the $\Gamma$XM plane and 60 eV for the ZRA plane were selected as shown in Fig.~\ref{figure1} (b). Figure~\ref{figure1} (c) and (d) display the results of the FS mapping for the photon energies of 88~eV and at 60~eV, respectively. These results were obtained by plotting the intensity within the energy window of $\pm$~20~meV from the $E_{\rm F}$ in the ARPES spectra. Comparing with the predicted FS which was produced by local-density-approximation (LDA) calculations \cite{Chikamatsu1, Hamada1, Livesay1}, we see a small electron pocket centered around the $\Gamma$ point and a large hole pocket centered around the A point. The dramatic changes in the FS with the photon energy, as well as the good agreement between the ARPES results and LDA calculations, indicate that our ARPES results do not reflect surface electronic structures but rather 3D bulk electronic structures with energy dispersion in the $k_\perp$ direction. The finite density of states inside the electron and hole FSs centered around the $\Gamma$ and A points, respectively, as well as the remanent FS in Fig.~\ref{figure1}~(c) centered around the M point, similar to the hole pocket around the A point in Fig.~\ref{figure1}~(d), are probably due to the effect of $k_\perp$ broadening \cite{Krempasky1, Krempasky2, Wadati1}.

\begin{figure}
\begin{center}
\includegraphics[width=0.7\linewidth]{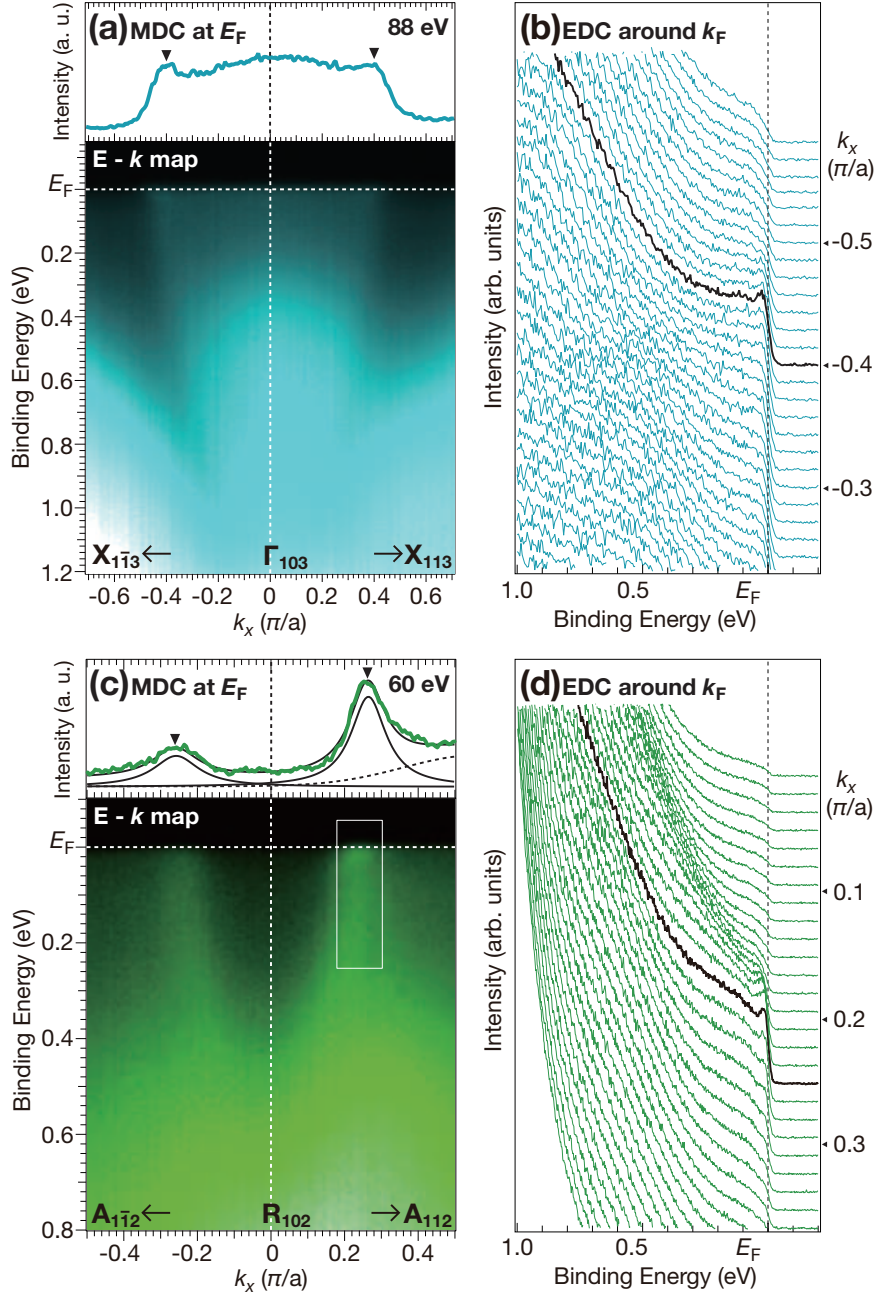}
\end{center}
\caption{(Color online) ARPES spectra of LSMO films taken at photon energies of 88~eV along the $\Gamma$-X direction [(a) and (b)] and of 60~eV along the R-A direction [(c) and (d)]. (a) and (c) Intensity map of ARPES spectra (lower panels) and MDCs at $E_{\rm F}$ (upper panels). Filled triangles indicate the $k_{\rm F}$ points determined by the peaks in the MDCs. The MDC along the R-A direction have been fitted to a linear combination of Lorentzians and a smooth background. (b) and (d) EDCs around the $k_{\rm F}$ points. Black thick spectra correspond to the EDCs at the $k_{\rm F}$ points.}
\label{figure2}
\end{figure}

\begin{figure}
\begin{center}
\includegraphics[width=0.8\linewidth]{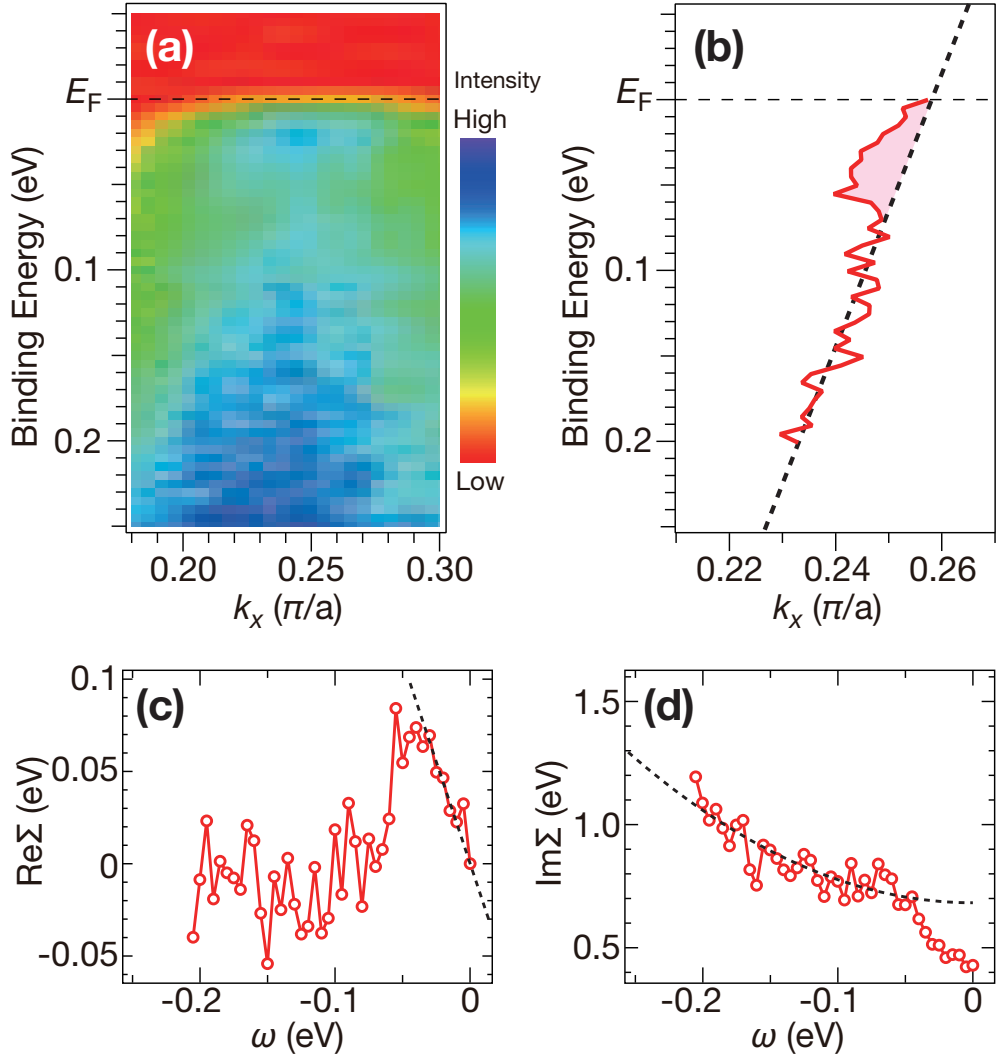}
\end{center}
\caption{(Color online) (a) Expanded intensity map for the hole band in the near-$E_{\rm F}$ and near-$k_{\rm F}$ regions [the box in Fig.~\ref{figure2} (c)]. (b) Band dispersion along the R-A direction determined by tracing the peak positions of MDCs. The dashed line indicates results of linear fitting on the band dispersion excluding the kink region. (c) Re$\Sigma(\omega)$ obtained from the MDC analysis of the ARPES spectra. The dashed line indicates the slope of Re$\Sigma(\omega)$ at $E_{\rm F}$, which corresponds to the coupling constant $\lambda$. (d) Im$\Sigma(\omega)$ obtained from the MDC analysis of the ARPES spectra. The dashed line indicates a fitted curve to the high-$\omega$ region of Im$\Sigma(\omega)$ with an $\omega^2$ function, which corresponds to the electron-electron interaction predicted by the Fermi-liquid theory.}
\label{figure3}
\end{figure}

The energy dispersion of the bands that form the FS is shown in Fig.~\ref{figure2}. In each FS, we observe a clear band dispersion with the Fermi cutoff, which reflects the metallic ground state of the LSMO. Pseudogap behaviors due to a nesting instability observed in 2D manganites \cite{Dessau} cannot be seen in the ARPES spectra of the 3D LSMO. In the energy distribution curves (EDCs) shown in Fig.~\ref{figure2} (b) and (d), one can see small but distinct fine peak structures near $E_{\rm F}$, which seem to have a "peak-dip-hump" structure, around the Fermi momentum ($k_{\rm F}$) in both the electron and hole bands. In order to examine the possible coupling of quasiparticle with collective excitations, we analyze the momentum distribution curves (MDCs) of the ARPES spectra. For the electron band, determining the exact MDC peak positions is difficult because of the substantial background due to the $k_\perp$ broadening from the 3D small electron pocket \cite{Chikamatsu1, Krempasky1, Krempasky2}. Therefore, we concentrate on analyzing the hole band. Figure~\ref{figure3} (a) shows an expanded plot of the intensity map in the near-$E_{\rm F}$ and near-$k_{\rm F}$ region for the hole band that was displayed in Fig.~\ref{figure2} (c). The intensity modulation that is derived from the peak-dip-hump structure is also exhibited in the intensity plot of the hole band. Figure~\ref{figure3} (b) displays the band dispersion obtained from the plot of the peak positions of the MDCs, which are determined by fitting the MDC to the linear combination of Lorentzians and a smooth background originating from the $k_\perp$ broadening of the hole pocket, which can be seen in the upper panel of Fig.~\ref{figure2} (c). A clear kink in the band dispersion can be observed at the binding energy of approximately 50~meV, which corresponds well to the peak-dip-hump structures in the EDCs.

In order to determine the characteristic energy scale for the observed kink structure, we have experimentally determined the real part [Re$\Sigma$($\omega$)] and the imaginary part [Im$\Sigma$($\omega$)] of the self-energy as functions of binding energy $\omega$ from the width $\Delta k$ and the peak positions $k_m$ of the MDCs in the ARPES spectra. If we assume that the bare band dispersion near $E_{\rm F}$ is linear, the Re$\Sigma$($\omega$) and Im$\Sigma$($\omega$) can be expressed by the following equations \cite{Valla3}:
\begin{eqnarray}
{\rm Re}\Sigma(\omega) &=& \omega - (k_m - k_{\rm F}) \cdot v_0 \nonumber \\
{\rm Im}\Sigma(\omega) &=& \Delta k \cdot v_0 /2
\label{equation1}
\end{eqnarray}
We can validate the approximations in the analysis of the LSMO because the experimental band dispersion can be fitted well using a linear function, excluding the kink region, as shown in Fig.~\ref{figure3} (b). The bare velocity $v_0$ is defined as the slope of this linear function, therefore the effects of electron-electron interactions and momentum dependence in Re$\Sigma$($\omega$) have been eliminated in this definition of Re$\Sigma$($\omega$). In addition, the hole band of the LSMO, which was predicted by the LDA calculations \cite{Chikamatsu1, Hamada1}, exhibits almost linear dispersion in the range of 200~meV from $E_{\rm F}$. Figure~\ref{figure3} (c) and (d) displays Re$\Sigma$($\omega$) and Im$\Sigma$($\omega$), respectively, which were obtained from the MDCs using these equations. The overall line shapes of both Re$\Sigma$($\omega$) and Im$\Sigma$($\omega$) are similar to those observed in the surface bands of the transition metals \cite{Valla2, Higashiguchi, Jiang}. This suggests that the origin of the kink in the ARPES band is an electron-boson coupling as in the case of the transition-metal surface bands. Re$\Sigma$($\omega$) exhibits a peak structure at the binding energy of approximately 50~meV, which corresponds to the kink in the ARPES band dispersion. It should be noted that the kink observed in the 50~-~60~meV energy region in this study is characteristic of another perovskite oxide SrVO$_3$ \cite{Aizaki}.

Coupling constant $\lambda~\sim~1.8$ of the electron-boson interaction is obtained from the slope of Re$\Sigma$($\omega$) at $E_{\rm F}$ \cite{Hayashi, Jiang}. The substantially large value of $\lambda$ compared with SrVO$_3$ indicates a much stronger coupling of the quasiparticle with boson excitations in the LSMO. The effective mass enhancement factor, $1~+~\lambda~\sim~2.8$, agrees well with the value deduced from the electronic specific heat coefficients \cite{Okuda}, which suggests that the enhanced effective mass in the LSMO could originate from a strong electron-boson coupling. Furthermore, Im$\Sigma$($\omega$) exhibits an $\omega^2$ dependence for binding energy values greater than 50~meV, which reflects the electron-electron interaction predicted by the Fermi-liquid theory. In the near-$E_{\rm F}$ region of within 50~meV from $E_{\rm F}$, Im$\Sigma$($\omega$) rapidly deceases as $\omega$ decreases. This also suggests the existence of the electron-boson interactions on the characteristic energy scale of $\sim$~50~meV.

\begin{figure}
\begin{center}
\includegraphics[width=0.8\linewidth]{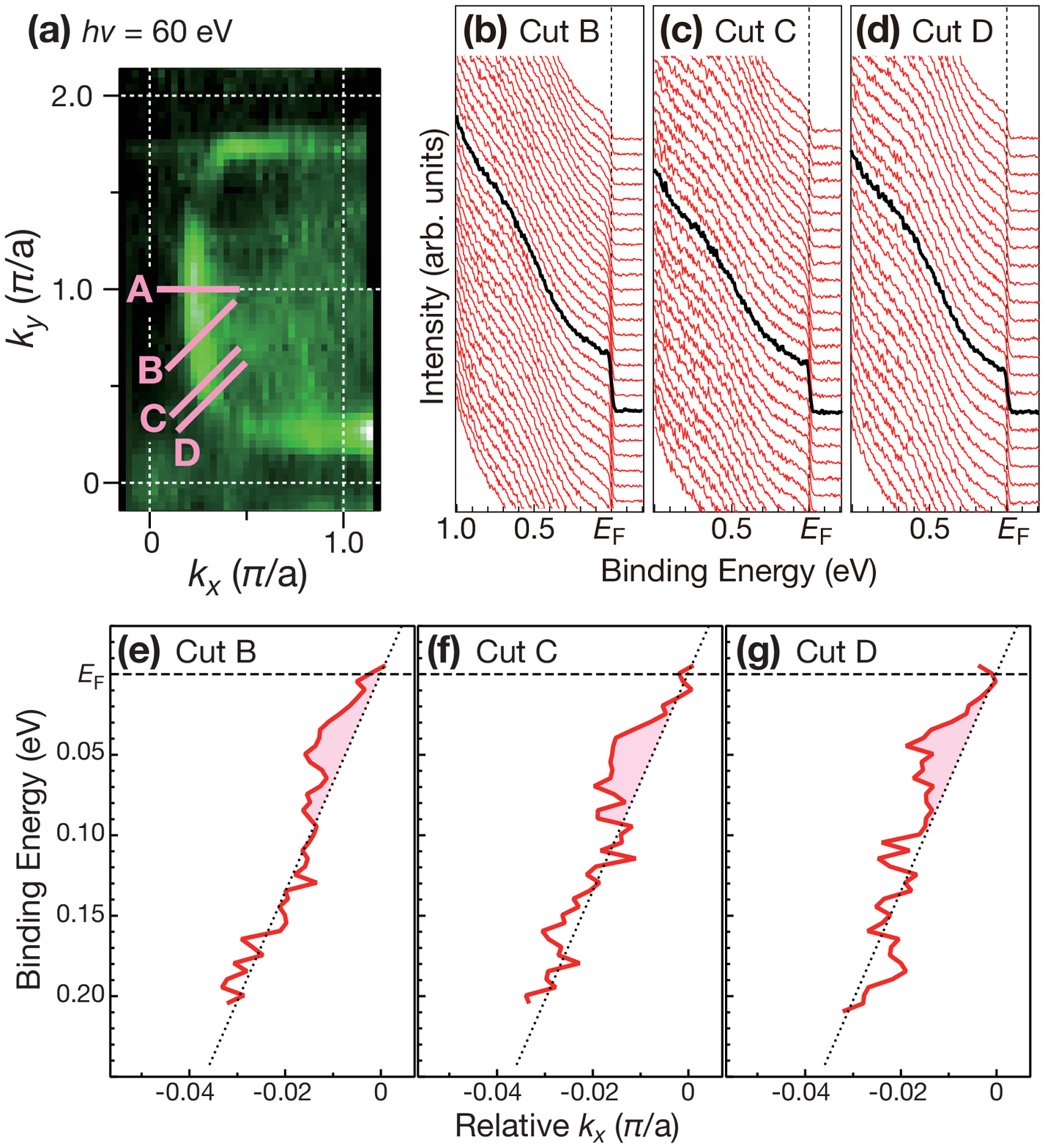}
\end{center}
\caption{(Color online) (a) FS mapping for photon energy of 60 eV corresponding to the ZRA plane, which is the magnification of Fig.~\ref{figure1} (d). (b) - (d) Momentum dependence of the EDCs around $k_{\rm F}$ and (e) - (g) of the band dispersion obtained from the MDCs along cuts B - D, respectively, in Fig.~\ref{figure4}~(a). Black thick spectra in EDCs correspond to the spectra at the $k_{\rm F}$ points. Note that the EDCs around $k_{\rm F}$ and the band dispersion along a cut A are shown in Fig.~\ref{figure2}~(d) and Fig.~\ref{figure3}~(b), respectively.}
\label{figure4}
\end{figure}

The quasiparticle peak with a kink has also been observed in the 2D layered manganite La$_{1.2}$Sr$_{1.8}$Mn$_2$O$_7$ \cite{Mannella1}. That kink has $\lambda$ as large as $\sim$~4.6 indicates the existence of an inherent strong electron-boson coupling in perovskite manganese oxides. However, the momentum dependence significantly differs between the 2D manganites and our 3D manganites. In La$_{1.2}$Sr$_{1.8}$Mn$_2$O$_7$, the quasiparticle peak is the sharpest along the (0, 0) to ($\pi$, $\pi$) diagonal directions, which is analogous to the high-$T_c$ cuprates. In contrast, our ARPES results for the 3D LSMO provide evidence for the existence of the kink and the sharp quasiparticle peak along the ($\pi$, 0, 0) to ($\pi$, $\pi$, 0) directions, i.e., the antinodal directions in the high-$T_c$ cuprates. Furthermore, the isotropic nature of the kink for the hole FS can be seen in Fig.~\ref{figure4}. Kinks in the band dispersion at almost the same energy of around 50~meV and sharp edges at $E_{\rm F}$ are observed throughout the hole FS from the antinodal to near-nodal directions without any signature of nesting instabilities as observed in 2D manganites \cite{Dessau, Mannella1}. Strongly localized bosons coupled with electrons most likely cause these isotropic kinks in the momentum space of 3D manganites.

We consider two plausible interactions in order to clarify the origin of the quasiparticle excitation and the resulting kinks in LSMO; the coupling of electrons with magnetic excitations and the coupling of electrons with phonon modes. Concerning the coupling of electrons with the magnetic excitations (namely, ferromagnetic magnons in double-exchange manganites), spin wave dispersions throughout the Brillouin zone have been investigated using the inelastic neutron scattering on La$_{0.7}$Pb$_{0.3}$MnO$_3$ \cite{Perring}. The magnon band is highly dispersive and has a bandwidth of $\sim$~100~meV, suggesting that electron-magnon interactions are broadened in the energy scale and limited to particular regions in momentum space, which are defined with critical momentum transfer vectors. Consequently, we expect the quasiparticle excitation and kink to have a large anisotropy in the momentum space. Moreover, the half-metallic electronic structure of LSMO does not allow magnon excitations to scatter electrons near $E_{\rm F}$ since spin-flip scattering cannot occur near $E_{\rm F}$ because of the full spin polarization.

As for the coupling with phonon modes, phonon dispersions have also been determined using inelastic neutron scattering on La$_{0.7}$Sr$_{0.3}$MnO$_3$ \cite{Reichardt, Zhang}. Several less-dispersive, that is, relatively localized phonon-mode branches are observed around the energies of 50~meV. In particular, a phonon branch, which is derived from the excitation of a Jahn-Teller (JT) phonon mode with linear-breathing characters, is located at the energies between $\sim$40~meV at the $\Gamma$ point and $\sim$60~meV at the zone boundary, which is close to the kink energy observed in the ARPES band dispersions. Therefore, it is reasonable to conclude that kink in the band dispersions results from the interaction of electrons with JT phonon modes. Strong interaction between electrons and local JT phonons in LSMO has been considered to lead to the formation of JT small polarons. However, the observation of the well-defined quasiparticles at least near $E_{\rm F}$ indicated that electrons (or holes) around $E_{\rm F}$ basically form energy bands and that the ideal small polaron formation is not very likely. Probably, polarons of intermediate size between small polarons and large polarons may be formed. JT distortions are collectively excited by electron (or hole) hopping between JT active Mn$^{3+}$ and inactive Mn$^{4+}$ ions. Consequently, the effective mass of quasiparticles is enhanced, even in the ferromagnetic metallic phase in LSMO. Indeed, a polaronic signature in ferromagnetic metallic manganites was recently observed in a scanning tunneling spectroscopic study of La$_{0.7}$Ca$_{0.3}$MnO$_3$ \cite{Seiro}.

In summary, we perform an \textit{in situ} ARPES study of 3D LSMO and find the isotropic quasiparticle excitation peaks and kinks in the experimental band dispersions. The observed isotropically strong renormalization of the band dispersions suggests that the electrons strongly interact with localized JT phonons and that the polaronic quasiparticles play an important role, even in the ferromagnetic metallic phase of LSMO.

\acknowledgements
The authors are very grateful to Prof. T.~Yoshida and Prof. K.~Ishizaka for fruitful discussions. This work was supported by a Grant-in-Aid for Scientific Research (B25287095 and S22224005), a Grant-in-Aid for Young Scientists (26870843) from the Japan Society for the Promotion of Science (JSPS), and the MEXT Elements Strategy Initiative to Form Core Research Center. M.~K. acknowledges the financial support from JSPS for Young Scientists. This work at KEK-PF was performed under the approval of the Program Advisory Committee (Proposals 2012G536, 2012G668, and 2013S2-002) at the Institute of Materials Structure Science, KEK.

\newpage

\begin{references}
\bibitem{CMR} \textit{Colossal Magnetoresistive Oxides}, edited by Y. Tokura, Advances in Condensed Matter Science, Vol. 2 (Gordon and Breach, Amsterdam, 2000).
\bibitem{MIT} M. Imada, A. Fujimori, and Y. Tokura, Rev. Mod. Phys. {\bf 70}, 1039 (1998).
\bibitem{Hufner1} S.~H\"{u}fner, \textit{Very High Resolution Photoelectron Spectroscopy} (Springer, New York, 2007).
\bibitem{Valla1} T.~Valla, A.~V.~Fedorov, P.~D.~Johnson, B.~O.~Wells, S.~L.~Hulbert, Q.~Li, G.~D.~Gu, and N.~Koshizuka, Science {\bf 285}, 2110 (1999).
\bibitem{Damascelli} A.~Damascelli, Z.~Hussain, and Z.-X.~Shen, Rev. Mod. Phys. {\bf 75}, 473 (2003).
\bibitem{Valla2} T.~Valla, A.~V.~Fedorov, P.~D.~Johnson, and S.~L.~Hulbert, Phys. Rev. Lett. {\bf 83}, 2085 (1999).
\bibitem{Schafer} J.~Sch\"{a}fer, D.~Schrupp, E.~Rotenberg, K.~Rossnagel, H.~Koh, P.~Blaha, and R.~Claessen, Phys. Rev. Lett. {\bf 92}, 097205 (2004).
\bibitem{Higashiguchi} M.~Higashiguchi, K.~Shimada, K.~Nishiura, X.-Y.~Cui, H.~Namatame, and M.~Taniguchi, Phys. Rev. B {\bf 72}, 214438 (2005).
\bibitem{Hayashi} H.~Hayashi, K.~Shimada, J.~Jiang, H.~Iwasawa, Y.~Aiura, T.~Oguchi, H.~Namatame, and M.~Taniguchi, Phys. Rev. B {\bf 87}, 035140 (2013).
\bibitem{Jiang} J.~Jiang, S.~S.~Tsirkin, K.~Shimada, H.~Iwasawa, M.~Arita, H.~Anzai, H.~Namatame, M.~Taniguchi, I.~Yu.~Sklyadneva, R.~Heid, K.-P.~Bohnen, P.~M.~Echenique, and E.~V.~Chulkov, Phys. Rev. B {\bf 89}, 085404 (2014).
\bibitem{Dessau} Y.-D.~Chuang, A.~D.~Gromko, D.~S.~Dessau, T.~Kimura, and Y.~Tokura, Science {\bf 292}, 1509 (2001).
\bibitem{Mannella1} N.~Mannella, W.~L.~Yang, X.~J.~Zhou, H.~Zheng, J.~F.~Mitchell, J.~Zaanen, T.~P.~Devereaux, N.~Nagaosa, Z.~Hussain, and Z.-X.~Shen, Nature {\bf 438}, 474 (2005).
\bibitem{Shi1} M.~Shi, M.~C.~Falub, P.~R.~Willmott, J.~Krempasky, R.~Herger, K.~Hricovini, and L.~Patthey, Phys. Rev. B {\bf 70}, 140407(R) (2004).
\bibitem{Falub1} M.~C.~Falub, M.~Shi, P.~R.~Willmott, J.~Krempasky, S.~G.~Chiuzbaian, K.~Hricovini, and L.~Patthey, Phys. Rev. B {\bf 72}, 054444 (2005).
\bibitem{Chikamatsu1} A.~Chikamatsu, H.~Wadati, H.~Kumigashira, M.~Oshima, A.~Fujimori, N.~Hamada, T.~Ohnishi, M.~Lippmaa, K.~Ono, M.~Kawasaki, and H.~Koinuma, Phys. Rev. B {\bf 73}, 195105 (2006).
\bibitem{Krempasky1} J.~Krempask\'{y}, V.~N.~Strocov, L.~Patthey, P.~R.~Willmott, R.~Herger, M.~Falub, P.~Blaha, M.~Hoesch, V.~Petrov, M.~C.~Richter, O.~Heckmann, and K.~Hricovini, Phys. Rev. B {\bf 77}, 165120 (2008).
\bibitem{Krempasky2} J.~Krempask\'{y}, V.~N.~Strocov, P.~Blaha, L.~Patthey, M.~Radovi\'{c}, M.~Falub, M.~Shi, and K.~Hricovini, J. Electron Spectrosc. Relat. Phenom. {\bf 181}, 63 (2010).
\bibitem{Tabeno} A.~Tebano, A.~Orsini, P.~G.~Medaglia, D.~Di~Castro, G.~Balestrino, B.~Freelon, A.~Bostwick, Y.~J.~Chang, G.~Gaines, E.~Rotenberg, and N.~L.~Saini, Phys. Rev. B {\bf 82}, 214407 (2010).
\bibitem{Horiba1} K.~Horiba, H.~Ohguchi, H.~Kumigashira, M.~Oshima, K.~Ono, N.~Nakagawa, M.~Lippmaa, M.~Kawasaki, and H.~Koinuma, Rev. Sci. Instrum. {\bf 74}, 3406 (2003).
\bibitem{Monkman1} E.~J.~Monkman, C.~Adamo, J.~A.~Mundy, D.~E.~Shai, J.~W.~Harter, D.~Shen, B.~Burganov, D.~A.~Muller, D.~G.~Schlom, and K.~M.~Shen, Nat. Mater. {\bf 11}, 855 (2012).
\bibitem{Horiba2} K.~Horiba, A.~Chikamatsu, H.~Kumigashira, M.~Oshima, N.~Nakagawa, M.~Lippmaa, K.~Ono, M.~Kawasaki, and H.~Koinuma, Phys. Rev. B {\bf 71}, 155420 (2005).
\bibitem{Hamada1} N.~Hamada \textit{et al}., unpublished.
\bibitem{Livesay1} E.~A.~Livesay, R.~N.~West, S.~B.~Dugdale, G.~Santi, and T.~Jarlborg, J. Phys.: Condens. Matter {\bf 11}, L279 (1999).
\bibitem{Wadati1} H.~Wadati, T.~Yoshida, A.~Chikamatsu, H.~Kumigashira, M.~Oshima, H.~Eisaki, Z.-X.~Shen, T.~Mizokawa, and A.~Fujimori, Phase Transitions {\bf 79}, 617 (2006).
\bibitem{Valla3} T.~Valla, T.~E.~Kidd, W.-G.~Yin, G.~D.~Gu, P.~D.~Johnson, Z.-H.~Pan, and A.~V.~Fedorov, Phys. Rev. Lett. {\bf 98}, 167003 (2007).
\bibitem{Aizaki} S.~Aizaki, T.~Yoshida, K.~Yoshimatsu, M.~Takizawa, M.~Minohara, S.~Ideta, A.~Fujimori, K.~Gupta, P.~Mahadevan, K.~Horiba, H.~Kumigashira, and M.~Oshima, Phys. Rev. Lett. {\bf 109}, 056401 (2012).
\bibitem{Okuda} T.~Okuda, A.~Asamitsu, Y.~Tomioka, T.~Kimura, Y.~Taguchi, and Y.~Tokura, Phys. Rev. Lett. {\bf 81}, 3203 (1998).
\bibitem{Perring} T.~G.~Perring, G.~Aeppli, S.~M.~Hayden, S.~A.~Carter, J.~P.~Remeika, and S.-W.~Cheong, Phys. Rev. Lett. {\bf 77}, 711 (1996).
\bibitem{Reichardt} W.~Reichardt and M.~Braden, Physica B  {\bf 263-264}, 416 (1999).
\bibitem{Zhang} J.~Zhang, P.~Dai, J.~A.~Fernandez-Baca, E.~W.~Plummer, Y.~Tomioka, and Y.~Tokura, Phys. Rev. Lett. {\bf 86}, 3823 (2001).
\bibitem{Seiro} S.~Seiro, Y.~Fasano, I.~Maggio-Aprile, E.~Koller, O.~Kuffer, and \O.~Fischer, Phys. Rev. B {\bf 77}, 020407 (R) (2008).
\end{references}
\end{document}